# A Parametric Chain based Routing Approach for Underwater Sensor Network

Aarti[#1], Sanjiv Kumar Tomar[#2]

[#1]Student, M.Tech,CSE,Amity University
Noida,India

[#2]Assistant Professor,CSE,Amity University
Noida,India

*Abstract:* A sensor network is one of the critical networks that is based on hardware components as well the energy parameters. Because of this, such network requires the optimization in all kind communication to improve the network life. In case of underwater sensor network, the criticality of network is also increased because of the random floating movement of the nodes. In this work, a composition of the multicast or broadcast communication is presented by the generation of aggregative path. The presented work is about to define a new chain based aggregative routing approach to provide the effective communication over the network. In this work, an effective aggregative path is suggested under the different parameters of energy, distance and congestion analysis. Based on these parameters a trustful aggregative route will be generated so that the network life will be improved.

## I. INTRODUCTION

Data aggregation[1] is one of the major communication approaches in which multiple sources are sending data to single sink. This communication approach is applicable to most of the network scenarios as each network is controlled by some controller node or the base station and all other nodes are agreed to report them regularly for different purpose such as for permissions, authentications, to connect to outer interface etc. In all such case we need to transfer data from multiple sources to the single sink. But as such kind of communication is performed individually between the each source and the sink. There will be heavy traffic or the communication over the network, that can result data and energy loss. Because of this there is the requirement of Data aggregation. The basic idea behind the aggregation is to avoid the one to one communication to the sink node and build an aggregative path over the network. The aggregative path is the path that selects a distance node as the source node and all the network nodes are intermediate nodes. The main objective of the data aggregation is to reduce the network traffic and to save the energy loss over the communication. While working with sensor network or any adhoc network, there are many dedicated applications that are suitable to the data aggregation.

The aggregation also defined under some constraints and the components. One of the such components is the aggregation operator called aggregator. The aggregator can be union operator that just combine the dataset taken from the previous node, add a delimiter and node message to it, and pass the message to next node. But such kind of aggregation is applicable only for the smaller network. As the network size grows, the union cannot be used. In such case, other aggregative operator called summation will be used. According to this operator the numerical values extracted from all nodes are added as the communication transferred to the next node. But these all kind of aggregation also suffer from the problem of inclusion of some false or bad information. In such case, the false data detection approach is used. The basic process of aggregation is shown in figure 1.

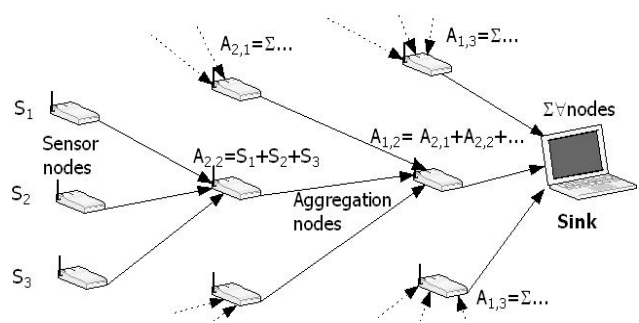

Figure 1: Principle of Data Aggregation [2]

In this figure a tree based aggregation is shown. In which all the end nodes transfer data directly to nearest cluster head, and the aggregation is performed between these cluster heads. The cluster head is responsible to use the aggregator. Here the summation is used for the aggregation. All these cluster-heads collectively generate an aggregative path and perform the communication to the sink node. The false data detection can be performed on each cluster head or on the sink node collectively.

The main aim of data aggregation is to reduce the energy consumption by generating effective aggregative path. The path selection in an aggregation depends on multiple factors such as type of network, network topology etc. In most cases, the decision regarding the aggregative path is been taken based on the type of network topology or the type of network. In such case, the network coverage cast is been analysed. One of such network construction type is the tree based network structure. In such formation, the aggregation is performed between the parent nodes and each parent collect data directly from the child nodes. In case of sensor network, the clustered architecture also gives the concept of aggregative communication between the cluster heads.

In a clustered network[3], a network is divided in smaller units called clusters and each cluster is controlled by the cluster head. These all cluster heads performs the communication to the base station. But if the cluster head is not in the sensing range of the base station, in such case cooperative communication between these cluster heads is performed and an aggregative path is generated between these cluster heads. Most of the sensor network protocols such as Leach[4] perform the aggregation on cluster heads. Whereas the protocols like PEGASIS[5] are the chain based protocols, that perform the aggregation between the nodes.

In this present work, the data aggregation scheme is defined for an underwater sensor network. The work is about to generate an aggregative path over the network that will connect all the network nodes in a chain so that the optimization will be obtained in terms of distance covered as well as in terms of energy consumed over the network. The aggregation is performed under the limits of underwater. According to these limits, the node localization is required to obtain the maximum level of connectivity between the nodes so that no node will be lost because of the absence of GPS. In this paper, the proposed aggregative path construction scheme is represented. In section 2, the existing work respective to the routing algorithm as well as the aggregation is defined by different authors. In section 3, the presented work proposal is shown and in section 5, the conclusion based on this work is shown.

## II.LITERATURE SURVEY

The presented work is about to generate an optimized route over the network so that all the network nodes will be covered. The work is about to present an effective aggregative path generation scheme under different parameters so that the energy and distance effective path will be generated. In this section, different routing approaches used by the earlier researchers is shown for any kind of sensor network.

M.R. Wilby et. al. [6], propose a new architecture called HARP, a Hierarchical Adaptive and Reliable Routing Protocol, a clustering algorithm which builds inter cluster and intra-cluster hierarchical trees, which are optimized to save power. The presented architecture is scalable and can be used in both homogeneous and heterogeneous wireless sensor networks. By means of the addition of a recovery slot in the scheduling scheme, HARP provides efficient link fault tolerance and also supports node mobility management. Young-Long et al. [7], propose the PEGASIS topology architecture with the PBCA (phase-based coverage algorithm) to find the redundant nodes which can enter to sleep mode. Therefore, our proposed algorithm can reduce the energy consumption of nodes and extend the network lifetime. Simulation results show that the performances of our algorithm out performance the LEACH topology architecture, the PEGASIS topology architecture, and the LEACH with PBCA topology architecture in terms of energy consumptions, number of nodes alive, and sensing areas. Feng Sen et. al. [8], propose EEPB (Energy-Efficient PEGASIS-Based protocol) is a chain-based protocol which has certain deficiencies including the uncertainty of threshold adopted when building a chain, the inevitability of long link (LL) when valuing threshold inappropriately and the non-optimal election of leader node. Aiming at these problems, an improved energy-efficient PEGASIS-based protocol (IEEPB) is proposed in this paper. IEEPB adopts new method to build chain, and uses weighting method when selecting the leader node, that is assigning each node a weight so as to represent its appropriate level of being a leader which considers residual energy of nodes and distance between a node and base station (BS) as key parameters. Wang Linping et. al. [9], proposed a new algorithm-PDCH, on the bases of PEGASIS to make every node load balance and extent the network lifetime. Protocol PEGASIS is based on the chain structure, every chain have only one cluster head, it is in charge with every node's receiving and sending messages who belong to this chain, the cluster head consumes large energy and the time of every round increases. Yongchang Yu et. al. [10], propose EECB (Energy-Efficient Chain-Based routing protocol) that is an improvement over PEGASIS. EECB uses distances between nodes and the BS and remaining energy levels of nodes to decide which node will be the leader that takes charge of transmitting data to the BS. Also, EECB adopts distance threshold to avoid formation of LL (Long Link) on the chain.

Young-Long Chen et. al. [11], proposes a model in order to reduce energy consumption. Author has first shown ideal energy mathematical model of PEGASIS topology, since the distance between nodes is the same, this energy mathematical model gives the longest network lifetime of WSNs. To achieve this objective, Author have proposed Intra- Grid PEGASIS topology architecture, which is an architecture based on PEGASIS topology; in this architecture, the sensor area is divided into several network grids, meanwhile, the nodes of each network grid is deployed in random, then the nodes within the network grid are connected, finally, all the network grids are connected. Wenjing Guo et. al. [12], proposes a routing protocol for the applications of Wireless Sensor Network (WSN). It is a protocol based on the PEGASIS protocol but using an improved ant colony algorithm rather than the greedy algorithm to construct the chain. Compared with the original PEGASIS, this one, Pegant, can achieve a global optimization. It forms a chain that makes the path more even-distributed and the total square of transmission distance much less. Moreover, in the constructing process, the energy factor has been taken into account, which brings about a balance of energy consumption between nodes. In each round of transmission, according to the current energy of each node, a leader is selected to directly communicate with the base station (BS). Simulation results have show that the proposed protocol significantly prolongs the network lifetime. M. Tabibzadeh et. al.[13], proposes a hybrid protocol, which author has called collectively Chain-based LEACH (CBL) that improves the Low-Energy Adaptive Clustering Hierarchy (LEACH) to significantly reduce energy consumption and increase the lifetime of a sensor network. The protocol uses LEACH and the advantages of Power-Efficient Gathering in Sensor Information Systems (PEGASIS) and avoids their disadvantages. LEACH technique improves energy efficiency of a sensor network by selecting a cluster-head, and having it aggregate data from other nodes in its cluster, and PEGASIS is a near optimal chain-based protocol that author uses for communication and extra aggregation between cluster-heads that are neighbors and takes turns transmitting to the sink. Hyunduk Kim et. al. [14], proposes DERP (Distance-based Energy-efficient Routing Protocol) is to address this problem. DERP is a chain-based protocol that improves the greedy-algorithm in PEGASIS by taking into account the distance from the HEAD to the sink node. The main idea of DERP is to adopt a pre-HEAD (P-HD) to distribute the energy load evenly among sensor nodes. In addition, to scale DERP to a large network, it can be extended to a multi-hop clustering protocol by selecting a "relay node" according to the distance between the P-HD and SINK. Analysis and simulation studies of DERP show that it consumes up to 80% less energy, and has less of a transmission delay compared to PEGASIS.

### III. RESEARCH METHODOLOGY

In this paper, a chain based aggregative routing is been presented for underwater sensor network. The work is defined in a single clustered network in which n number of nodes are defined with specific parameters such as energy, position, load, congestion vector etc. A base station is also defined at one end of the network. The work is about to define an optimized path that will be started from the farthest node from base station and after covering all the network nodes, transfer the aggregative data to the base station. The basic assumptions we have taken about the network are given as under

*A. Assumptions*

   a) All the nodes are in floating motion because the sensor network type is underwater
   b) The base station is placed at one end of the network.
   c) Absence of the GPS so that tracking of the nodes' location is maintained by the base station.
   d) To maintain the localization information a periodic energy is consumed called adaptive energy.

   e) All the sensor nodes are homogenous but they are having different failure factor, congestion factor and the energy level.

*B. Work Description*

The main idea for each node to receive from and transmit to close neighbours and take turns being the leader for transmission to the BS. The nodes will be organized to form chain which can either be accomplished by sensor nodes themselves using greedy algorithm starting from some node , the BS can compute this chain and broadcast it to all sensor nodes . When a node dies , the chain is reconstructed in the same manner to bypass the dead node. Nodes take turns transmitting to the BS and we will use node number I mod N (N represents the number of nodes) to transmit to the BS in round i . The leader in each round of communication will be at a random position on the chain,

A simple control token passing approach initiated by leader to start the data transmission from the ends of the chain. Each node will fuse its neighbours data with its own to generate a single packet of the same length and then transmit that to its other neighbour Each node will receive and transmit one packet in each round and be the leader once every 100 rounds.

The basic methodology respective to the defined parameters is given as under

1. While Selection of Next Node find
   a. Distance*Ei depending on Range
   b. Check Energy-Distance*Ei >Threshold
   c. Congestion Analysis
   d. Node Failure Analysis
   e. Check For Loop Back.
2. If Look Back(True)
   a. Then
      1. Discard that node
3. If Energy-distance*Ei <Threshold
   a. Then
      1. Discard that Node
4. Find the Node having
   i. Minimum Energy Consumption respective to Distance
   ii. Will Have Max Energy After Transmission
   iii. Minimum Load
   iv. Minimum Failure Probability
   v. Effective Throughput
   vi. Not performing the loop.
5. Perform the Communication

*C. Algorithm*

The proposed algorithm to present the proposed work is given as under

1) Define a network with N number of Nodes with specification of different parameters like energy, congestion vector etc.
2) Represent the Base Station as Destination Node
3) Find the farthest node from the base station and identify it as the source node.
4) Generate a localization table to track the maximum connectivity between the nodes.
5) Set S as source node and perform the communication
6) While S<> DestNode
7) [Repeat steps 7 to ]
8) Build the list of neighbouring nodes to Node S called Ne1,Ne2…Nem
9) For i=1 to M
10) {
11) Find Different parameters for each neighbouring nodes called distance energy, residual energy, congestion etc.
12) If(Energ(Neighbor(i))=MinEnergy and ResidualEnerg(Neighbor(i))=MaxResidualEnergy and Distance(i)=MinimumDist) And congestion(Neighbor(i))=Minimum
13) {
14) Set Neighbor(i) as Next Node
15) }
16) Else if (Energ(Neighbor(i))=MinEnergy and ResidualEnerg(Neighbor(i))=MaxResidualEnergy)
17) {
18) Set Neighbor(i) as Next Node
19) }
20) Else if (Energ(Neighbor(i))=MinEnergy and Distance(Neighbor(i))=MinDistance)
21) {
22) Set Neighbor(i) as Next Node
23) }
24) Else if (Energ(Neighbor(i))=MinEnergy)
25) {
26) Set Neighbor(i) as Next Node
27) }
 }
28) Set Si=SelectedNeighbour and repeat step 3 to 6 till Di not found }

IV. CONCLUSION

The presented work is about to generate an effective aggregative path over the network under multiple parameter analysis. The parameters included in this proposed work are distance, energy, residual energy and the congestion. The work is about to generate a chain based path over a less congested nodes so that network life will be improved. In this paper, the algorithmic approach for next neighbour selection is defined.